\documentclass[aps,showpacs,nofootinbib,eqsecnum,prd,twocolumn,superscriptaddress]{revtex4-1}
\usepackage{feynmp,slashed,color,mathrsfs,nicefrac,exscale}
\usepackage[colorlinks=true,linkcolor=blue,citecolor=blue,urlcolor=blue]{hyperref}

\usepackage{epsfig}
\usepackage[colorlinks=true,linkcolor=blue,citecolor=blue,urlcolor=blue]{hyperref}

\usepackage{amsmath}
\usepackage{amssymb}
\usepackage{marginnote}
\usepackage{braket}
\usepackage{xcolor}
\usepackage{mathtools}

\usepackage{bm,longtable,enumerate}

\def\beq{\begin{equation}}
\def\eeq{\end{equation}}
\def\beqa{\begin{eqnarray}}
\def\eeqa{\end{eqnarray}}
\def\bea{\begin{eqnarray}}
\def\eea{\end{eqnarray}}

\begin{document}

\title{An effective holographic approach to QCD}

\author{Alfonso Ballon-Bayona} 
\thanks{Invited talk presented at the XIV International Workshop on Hadron Physics, Florian\' opolis, Brazil, March 2018.} 
\affiliation{Instituto de F\'{\i}sica Te\'orica, Universidade Estadual Paulista,  Rua Dr. Bento Teobaldo Ferraz, 271 - Bloco II,
  01140-070 S\~ao Paulo, SP, Brazil}

\author{Henrique Boschi-Filho} 
\affiliation{Instituto de F\'{i}sica, Universidade
Federal do Rio de Janeiro,\\
Caixa Postal 68528, RJ 21941-972, Brazil}

\author{Luis A. H. Mamani} 
\affiliation{Centro de Ci\^encias Naturais e Humanas,
Universidade Federal do ABC,\\
Rua Santa Ad\'elia 166, 09210-170, Santo Andr\'e, SP, Brazil}

\author{Alex S. Miranda}  
\affiliation{Laborat\'orio de Astrof\'{\i}sica Te\'orica e Observacional\\ 
Departamento de Ci\^encias Exatas e Tecnol\'ogicas\\
Universidade Estadual de Santa Cruz, 45650-000, Ilh\'eus, BA, Brazil}

\author{Vilson T. Zanchin}  
\affiliation{Centro de Ci\^encias Naturais e Humanas,
Universidade Federal do ABC,\\
Rua Santa Ad\'elia 166, 09210-170, Santo Andr\'e, SP, Brazil}


\begin{abstract}

We describe a holographic approach to QCD where conformal symmetry is broken explicitly in the UV by a relevant operator ${\cal O}$. The operator maps to a 5d scalar field, the dilaton, with
a massive term. Implementing also the IR constraint found by Gursoy, Kiritsis and Nitti, an approximate linear glueball spectrum is obtained which is consistent with lattice data. Finally, we describe
the evolution of the model parameters with the conformal dimension of ${\cal O}$. This suggests a map between the QCD trace anomaly and the trace Ward identity of deformed conformal field theories.

\end{abstract}

\maketitle

\section{Introduction}
\label{sec:Introd}

In this proceedings, we describe the holographic model proposed in \cite{Ballon-Bayona:2017sxa}, where large-$N_c$ QCD is approximated by an effective theory arising from a CFT deformation.  We start in section \ref{sec:Introd} with an introduction  to the holographic QCD approach. In section \ref{sec:EffHQCD} we describe the main results of our model and we finish with our conclusions. 

\vspace{-0.4cm}

\subsection{Traditional non-perturbative approaches to  QCD}
\label{subsec:nonpertQCD}

QCD is the theory of strong interactions. 
The QCD lagrangian enjoys a non-Abelian gauge symmetry $SU(N_c)$ with $N_c=3$ is the number of colors. The fundamental degrees of freedom in QCD are the quarks and  gluons. The quarks are described by Dirac spinors $\psi_f$ in the fundamental representation of $SU(N_c)$ whereas the gluons are described by a non-Abelian gauge field $A_\mu$ in the adjoint representation of $SU(N_c)$. Despite the simplicity of the QCD lagrangian, QCD is a remarkably difficult theory in the regime of hadronic interactions. The reason is that at the quantum level the  coupling $g$ increases when lowering the energy scale $\mu$. This is described by the beta function $\beta_g = \mu \partial_\mu g$ which turns out to be negative for QCD. In the regime of typical hadronic interactions the coupling $g$ is so strong that perturbative methods are not reliable. 

The traditional non-perturbative approaches to QCD are lattice QCD and the Dyson-Schwinger (DS) equations. Lattice QCD deals with a discretized version of the QCD lagrangian that permits numerical simulations. It is an efficient approach for calculating the hadronic spectrum and thermodynamic properties of the quark-gluon plasma. The limitation of this approach regards real-time dynamics, due to the restriction to an Euclidean space.
The DS equations consists of a set of equations for quantum field theory correlators. In QCD, those equations can establish a bridge between the quark/gluon description and the hadronic dynamics. The difficulty in this approach regards the truncation of Feynman diagrams. Other non-perturbative approaches to QCD include chiral lagrangians, Nambu-Jona-Lasinio models, quark-meson models, QCD sum rules and renormalization group (RG) approaches. Finally, an alternative approach, proposed by 't Hooft in 1974, consists of taking the limit of large $N_c$ so that $1/N_c$ becomes a perturbative parameter. In this limit, only Feynman diagrams with planar topology survive. The obvious limitation of this approach is the fact that in QCD we have $N_c=3$, which is not large. This means that non-planar diagrams are generally required. An interesting feature of the $1/N_c$ expansion is that the Feynman diagrams are classified in terms of Riemann surfaces with different genus. The latter appear naturally in string theory. 

\vspace{-0.5cm}

\subsection{AdS/CFT and holographic QCD}
\label{subsec:CFTtoQCD}

The pure glue sector of the QCD lagrangian is  known as the Yang-Mills theory. Interestingly, at the classical level the Yang-Mills theory is invariant under the conformal group $SO(2,4)$, which is an extension of the Poincar\'e group $SO(1,3)$. Besides the ordinary Poincar\'e transformations (translations, rotations and boosts), the conformal group also includes the dilation (scale transformation) and special conformal transformations (inversion-translations-inversion). 

Remarkably, there is a supersymmetric extension of the Yang-Mills theory that preserves conformal symmetry at the quantum level. This is the so called ${\cal N}=4$ super Yang-Mills (SYM) theory  and  it is constrained by conformal symmetry in a dramatic fashion. In particular, the beta function of the theory is zero. The ${\cal N}=4$ SYM theory is one of the paramount examples of conformal field theories (CFT). At weak coupling, the theory also arises from string theory when considering a set of 4d hypersurfaces called D3-branes. In 1997, Maldacena realized that ${\cal N}=4$ SYM theory in $\mathbb{R}^{1,3}$ actually has a string theory dual (known as type IIB) that lives in a 10d spacetime, namely $AdS_5 \times S^5$. This is an example of the so called AdS/CFT correspondence, a correspondence between theories with gravity in anti-de Sitter space (AdS) and conformal field theories (CFT) without gravity in a lower dimension. The AdS/CFT correspondence is considered a realization of the holographic principle for a quantum theory of gravity.

\begin{figure}[h]
\centering
\includegraphics[width=8.75cm]{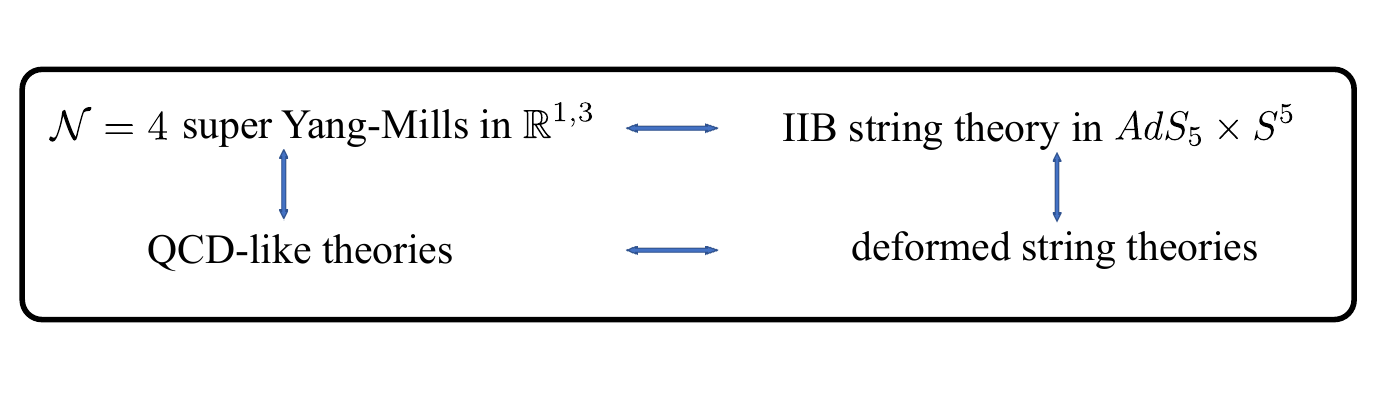} 
\vspace{-1cm}
\caption{The top-down approach} \label{Fig:TopDown}
\end{figure}

\begin{figure}[h]
\centering
\includegraphics[width=8.75cm]{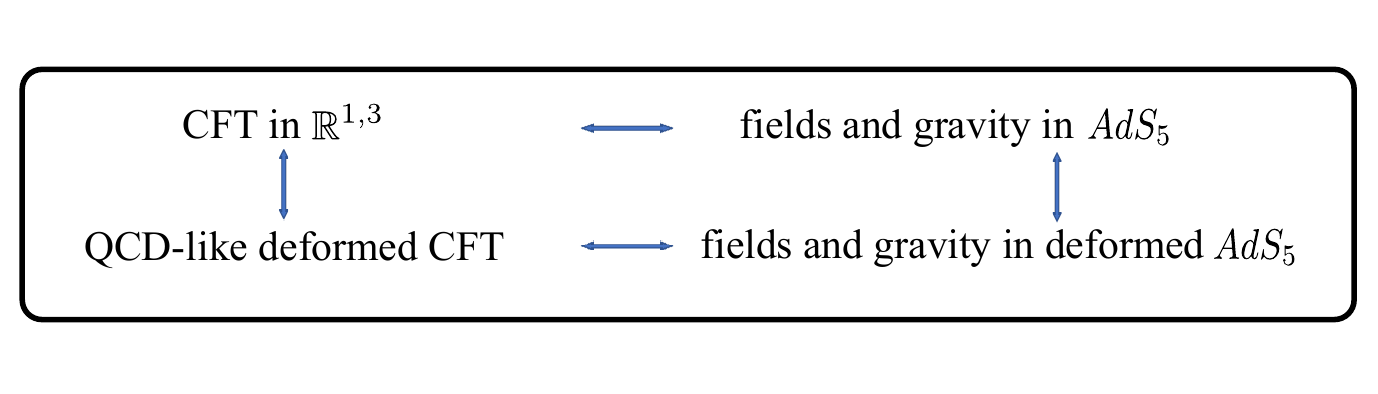} 
\vspace{-1cm}
\caption{The bottom-up approach} \label{Fig:BottomUp}
\end{figure}

The AdS/CFT correspondence motivated a research program known as holographic QCD, or AdS/QCD. The long term goal of holographic QCD is to find the 5d string theory (or quantum gravity) dual of QCD. There are two complementary approaches in holographic QCD, dubbed top-down and bottom-up. The top-down approach considers deformations of the original brane setup that led to the AdS/CFT correspondence. At weak coupling, one finds quantum field theories where conformal symmetry (and some supersymmetry) is broken. At strong coupling, one finds string theories in backgrounds that are less symmetric than the original $AdS_5 \times S^5$. Investigating those string theories one is able to make predictions for the strongly coupled regime of the dual QCD-like theories.  The top-down logic is depicted in Fig. \ref{Fig:TopDown}. The bottom-up approach, on the other hand, does not rely on string theory. In its original version, the bottom-up approach was considered  purely phenomenological in the sense that the goal was to incorporate QCD features by introducing a minimal set of fields in $AdS_5$. However, the modern bottom-up perspective is the construction of holographic models for the RG flow of deformed CFTs that are similar to QCD. Extending the traditional AdS/CFT dictionary between  operators and fields one is able to map 4d deformed CFTs to 5d theories with gravity living in a deformed $AdS$ background. The bottom-up logic is depicted in Fig. \ref{Fig:BottomUp}. For more details on the top-down and bottom-up approaches in holographic QCD, there are some nice reviews available \cite{Edelstein:2006kw,Erdmenger:2007cm,Erlich:2009me,Gursoy:2010fj,Brodsky:2014yha,Gursoy:2016ebw}. 

\vspace{-0.5cm}

\subsection{Holographic QCD from dilaton-gravity}
\label{subsec:HQCD}

A well known feature of (massless) QCD is the so called trace anomaly, associated with conformal symmetry breaking. In the large-$N_c$ limit, the trace anomaly equation can be written as 
\beqa
- \langle T^{\mu}_{\, \, \mu} \rangle = \frac{\beta_\lambda}{2 \lambda^2} \langle {\rm Tr} F^2 \rangle \, , \label{QCDTraceAnom}
\eeqa
where $\lambda=g^2 N_c$ is the 't Hooft coupling, $T^{\mu}_{\, \, \mu}$ is the trace of the energy-momentum tensor and ${\rm Tr} F^2$ is the Yang-Mills operator associated with the gluon condensate \footnote{The trace is defined as $\eta_{\mu \nu} T^{\mu \nu}$  and we consider here the signature $(-,+,+,+)$ for the Minkowski metric $\eta_{\mu \nu}$.}. Eq. \eqref{QCDTraceAnom} is exact at all loops and the two sides are RG invariant. 

Motivated by the QCD trace anomaly, in the bottom-up approach one could start building a minimal holographic model for the QCD vacuum. The minimal model contains only the 5d fields dual to the 4d operators associated with the QCD vacuum, namely the energy-momentum tensor $T^{\mu \nu}$ and Yang-Mills operator ${\rm Tr} F^2$. The dual fields are the 5d metric $g_{mn}$ and the 5d scalar field $\Phi$. The latter is usually called the dilaton, a field that arises naturally in string theory. This field is the responsible for the deformation of $AdS$ spacetime. 

The simplest action for describing the dynamics of $g_{mn}$ and $\Phi$ is a 5d dilaton-gravity action which can be written as \footnote{The action is defined in the Einstein frame.}
\beqa
\hspace{-0.8cm}
S \! = M_p^3 N_c^2 \int d^5 x \sqrt{-g} \Big [ R - \frac43 g^{mn} \partial_m \Phi \partial_n \Phi + V(\Phi) \Big ] , 
\eeqa
where $M_p$ is the 5d Planck mass scale. Varying this action we obtain the dilaton-gravity equations
\begin{align}
R_{mn} &= \frac43 \partial_m \Phi \partial_n \Phi - \frac13 g_{mn} V\, , \nonumber \\ 
\frac43 \nabla^2 \Phi + \frac12 \frac{d V}{d \Phi} &= 0 \,, 
\label{GenDilGravEqs}
\end{align}
In holographic QCD we take a general Poincar\'e-invariant ansatz 
\beqa
ds^2 = e^{2A(z)} \left [ - dt^2 + dx_i^2 + dz^2 \right ] \, , \,
\Phi = \Phi(z) . 
\eeqa
The dilaton-gravity equations then reduce to 
\begin{align}
A'^2 - A'' &= \frac49 \Phi'^2 \, \,  , \, \, 
3A'^2 + A'' = \frac{V}{3} e^{2A}  \,, \nonumber \\
\frac83  \left [ \partial_z + 3 A' \right ] \Phi' &= - e^{2A} \frac{dV}{d\Phi} \, ,
 \label{DilGravEqs}
\end{align}
where $'$ means $d/dz$. The last equation in \eqref{DilGravEqs} is not independent (it can be obtained from the first two).  Introducing the function $\zeta(z)\equiv \exp [ -A(z)]$, the first equation in \eqref{DilGravEqs} takes the form
\beqa
\zeta'' - \frac49 \Phi'^2 \zeta = 0 \,. 
\eeqa
In this way we easily recover the AdS spacetime $\zeta_0(z) = z/\ell$ when $\Phi$ is constant. A remarkable feature about the 2\textsuperscript{nd} order differential equations \eqref{DilGravEqs} is the fact that they can be brought into the following set of 1\textsuperscript{st} order differential equations
\begin{align}
\zeta' &= \frac49 W \quad  , \quad  \zeta \Phi' = \frac{dW}{d\Phi} \, , \nonumber \\
V &= \frac{64}{27} W^2 - \frac43 \frac{dW}{d \Phi} \,.
\end{align}
The functional $W[\Phi]$ is known as the superpotential and has a key role in the holographic description of the RG flow. Another important quantity is the ``holographic beta function'', defined as 
\beqa
\beta_\Phi \equiv \frac{d \Phi}{d A} = - \frac94 \frac{ d \log W}{d \Phi} \,. \label{hologbeta}
\eeqa
The improved holographic QCD models (IHQCD), proposed by Gursoy, Kiritsis and Nitti in \cite{Gursoy:2007cb,Gursoy:2007er} (see also \cite{Gursoy:2016ebw,Gursoy:2010fj}), relate the 't Hooft coupling $\lambda$ to the dilaton field $\Phi$ by $\lambda = e^{\Phi}$. Moreover, they consider also the traditional IR/UV correspondence $A(z) = \log \mu$ between the warp factor $A(z)$ and the RG energy scale $\mu$, so that the function in \eqref{hologbeta} becomes 
\beqa
\beta_\Phi = \frac{\beta_\lambda}{\lambda}  \,. \label{betafnrel}
\eeqa
This leads to a UV constraint (small $z$) for the background fields $A(z)$ and $\Phi(z)$. They should be compatible with the large-$N_c$ beta function of perturbative QCD. At two loops it is given by 
\beqa
\beta_\lambda = - b_0 \lambda^2 - b_1 \lambda^3 \, , \label{betapQCD}
\eeqa
where $b_0$ and $b_1$ are constants. The corresponding UV potential can be written as 
\beqa
V(\Phi) = \frac{12}{\ell^2} \left [ 1 + v_0 e^{\Phi} + v_1 e^{2 \Phi} + \cdots \right ] \, ,
\eeqa
with $v_0$ and $v_1$ determined from $b_0$ and $b_1$. A nice feature of holographic QCD is that the IR is naturally constrained by confinement and the hadronic spectrum. For the case at hand, it was found in \cite{Gursoy:2007cb,Gursoy:2007er} that the only IR behaviour for the dilaton, compatible with confinement and an asymptotic linear glueball spectrum, is 
\beqa
\Phi(z) = C z^2 + \cdots \, . 
\label{IRPhi}
\eeqa
The corresponding IR potential takes the form 
\beqa
V(\Phi) \sim  \Phi^{1/2} \exp \left[ \frac43 \Phi \right] \,. 
\label{IRPot}
\eeqa
In the IHQCD approach of \cite{Gursoy:2007cb,Gursoy:2007er}, the RG flow of the 4d theory starts already in the extreme UV ($\mu \to \infty$) where the theory is conformal and free. The 4d theory then is interpreted as the closest description of large-$N_c$ QCD in the sense that it is compatible with asymptotic freedom in the UV and confinement in the IR. In the next section we describe an alternative approach where QCD is approximated by an effective theory instead. The effective theory has a UV cutoff $\mu^*$ and we call this  approach effective holographic QCD (EHQCD). It was originally proposed in \cite{Gubser:2008ny,Gubser:2008yx} to describe the physics of the quark-gluon plasma.  In \cite{Ballon-Bayona:2017sxa}, we combine some features of  \cite{Gursoy:2007cb,Gursoy:2007er} and \cite{Gubser:2008ny,Gubser:2008yx} in order to arrive at a realistic description of the glueball spectrum. The results in \cite{Ballon-Bayona:2017sxa} also bring new insights into the holographic dictionary for this class of models. 

\section{The effective holographic approach to QCD}
\label{sec:EffHQCD}

\begin{figure}[tbp]
\centering
\includegraphics[width=8.75cm]{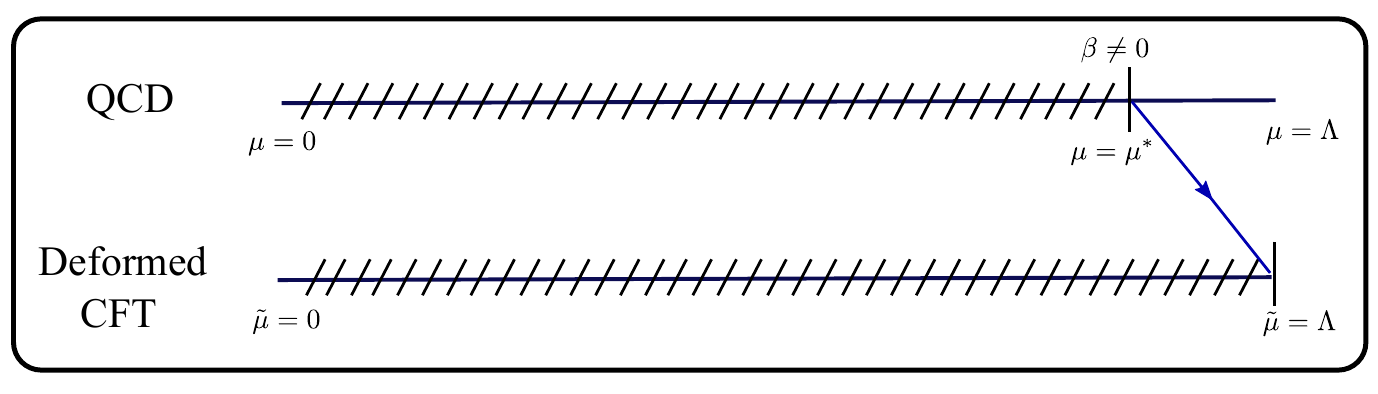} 
\vspace{-0.5cm}
\caption{A possible map between QCD and a deformed CFT. The extreme UV cutoff $\Lambda$ is taken to infinity in both cases.} \label{Fig:EHQCD}
\end{figure}

As explained in the previous section, conformal symmetry breaking in (massless) QCD is associated with the trace anomaly \eqref{QCDTraceAnom}. This is a quantum effect due to the renormalization of the theory. An alternative mechanism for conformal symmetry breaking is the following. Consider the deformation of a 4d CFT by a relevant scalar operator ${\cal O}$ with conformal dimension $\Delta = 4 - \epsilon$. The lagrangian can be written as 
\beqa
{\cal L} = {\cal L}_{CFT} + \phi_0 {\cal O} \, , \label{DefCFT}
\eeqa
where $\phi_0$ is the coupling associated with the deformation. The conformal dimension of $\phi_0$ is $4-\Delta=\epsilon$. The deformed CFT in \eqref{DefCFT} satisfies the trace Ward identity 
\beqa
- \langle T^{\mu}_{\, \, \mu} \rangle = (\Delta - 4) \phi_0 \langle {\cal O} \rangle \,. \label{traceDefCFT}
\eeqa
The similarity between \eqref{QCDTraceAnom} and \eqref{traceDefCFT} suggests that there might be a way of reformulating QCD in terms of a deformed CFT. In the perturbative regime of QCD, this means that after renormalization at one or two loops the effective action for QCD could be recast as a relevant deformation of a CFT. If that was the case the map between these two theories would look schematically like Fig. \ref{Fig:EHQCD}. One possible strategy to finding this map is considering QCD in $d=4-\epsilon$ so that the Yang-Mills operator ${\rm Tr} F^2$ becomes relevant. Interestingly, there has been recent progress establishing connections between quantum field theories in $d=4-\epsilon$ at the Wilson-Fisher fixed point and CFTs, see e.g. \cite{Rychkov:2015naa}.

\vspace{-0.5cm}

\subsection{The 5d background} 
\label{subsec:backg}

According to the AdS/CFT correspondence, a scalar operator ${\cal O}$ with conformal dimension $\Delta$ corresponds to a 5d scalar field $\Phi$ with a squared mass $M^2  = \ell^{-2} \Delta (\Delta - 4)$. In holographic QCD we then impose the following UV behaviour for the potential 
\beqa
V(\Phi) = \frac{12}{\ell^2} - \frac43 M^2 \Phi^2 + \cdots \, .
\label{UVPot}
\eeqa
Plugging this potential in Eqs. \eqref{DilGravEqs} one finds that the constant term leads to AdS asymptotics and the mass term leads to the dilaton asymptotics 
\beqa
\Phi(z) = \phi_0 z^{\Delta_{-}} + G z^{\Delta_{+}} \, , 
\label{UVPhi}
\eeqa
with $\Delta_{+} = \Delta = 4 - \epsilon$ and $\Delta_{-} = 4 - \Delta = \epsilon$. The coefficient $\phi_0$ is the source shown in \eqref{DefCFT} and the coefficient $G$ is proportional to the VEV of ${\cal O}$. The dilaton asymptotics is in turn responsible for the AdS deformation
\begin{align}
\label{UVWarpFactor}
\! \! A(z) &=
-\log{( z/ \ell )}-
\frac{2\Delta_{-} \phi_{0}^{2}}{9(1+2\Delta_{-})} z^{2\Delta_{-}}  
-
\frac{2\Delta_{-}\Delta_{+} \phi_{0}G}{45} z^4 \nonumber \\
&-
\frac{2 \Delta_{+} G^{2}}{9(1+2\Delta_{+})} z^{2\Delta_{+}}
-\cdots\,.
\end{align}
The UV behaviour for the holographic beta function \eqref{hologbeta}  is given by
\beqa
\beta_{\Phi} = - \Delta_{-} \phi_0 \,  z^{\Delta_{-}} - \Delta_{+} G \, z^{\Delta_{+}}  + \cdots \, . \label{UVhologbeta}
\eeqa
There is an important difference between the EHQCD and IHQCD models regarding the function in \eqref{UVhologbeta}. Although this function is associated with the RG flow of the dual theory, it is not proportional to the large-$N_c$ QCD beta function $\beta_\lambda$, as in \eqref{betafnrel}. In the EHQCD approach the dual theory is an effective theory for QCD with beta function $\tilde \beta_{\tilde \lambda}$. We will find later a dictionary for the original $\beta_\lambda$ from considerations of the trace anomaly \eqref{QCDTraceAnom}.  
Although the UV behaviour of the EHQCD models and IHQCD models is different, both provide a good description of confinement and the glueball spectrum as long as the IR behaviour is the same. In \cite{Ballon-Bayona:2017sxa} we proposed models that interpolate between the UV behaviour \eqref{UVPot}-\eqref{UVPhi} and the IR behaviour \eqref{IRPhi}-\eqref{IRPot}.  The simplest models are obtained by considering an analytic form for the dilaton and solving the warp factor numerically. Alternatively, one can also start with an analytic form for the warp factor and find a numerical solution for the dilaton. We call these two type of models A and B respectively. 

One simple interpolation for the dilaton is the following:
\beqa \label{interp1}
\Phi (z)= \hat \phi_0 (\Lambda z)^{\epsilon} 
+\frac{(\Lambda z)^{4-\epsilon}}{1+(\Lambda z)^{2-\epsilon}}\,,
\eeqa
where $\hat \phi_0$ and $\Lambda$ are related to $\phi_0$  and $G$ by
\beqa
\phi_0 = \hat \phi_0 \Lambda^{\epsilon} \quad , \quad 
G = \Lambda^{4-\epsilon} \,. 
\eeqa
We call this model A1. The coefficient $\hat \phi_0$ is the dimensionless version of the coupling and the coefficient $\Lambda$ behaves as an auxiliary field of conformal dimension $1$. The warp factor is solved numerically from Eqs. \eqref{DilGravEqs} imposing the UV condition \eqref{UVWarpFactor}. 

\begin{figure}[tbp]
\centering
\includegraphics[width=5.2cm,angle=0]{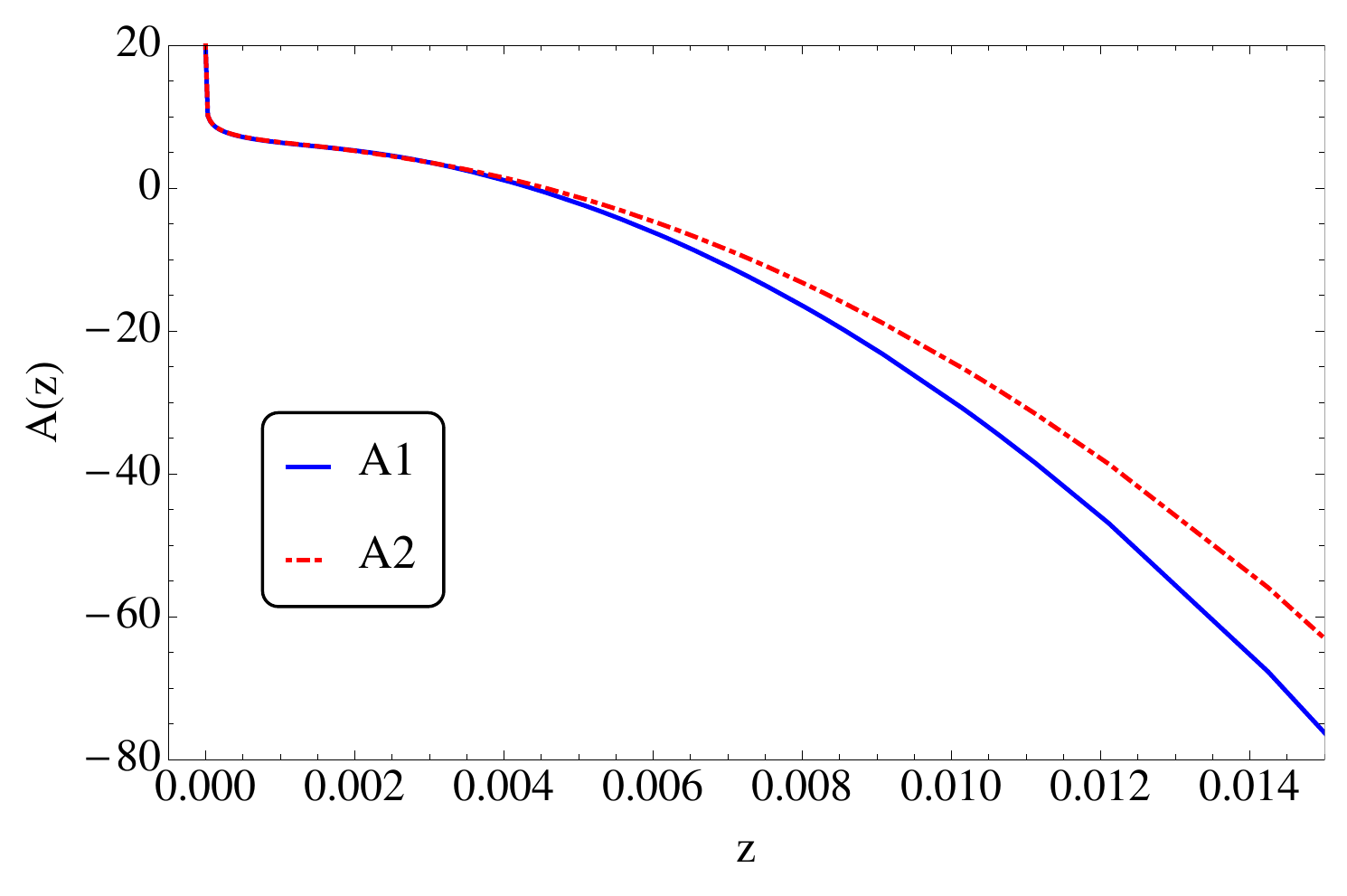} 
\vspace{-0.3cm}
\caption{Warp factor  $A(z)$ for models A1 and A2, obtained by solving numerically the dilaton-gravity equations \eqref{DilGravEqs}.}
\label{Fig:WarpFactor}
\end{figure}

Another simple interpolation can be built in terms of the hyperbolic tangent function, 
\beqa\label{interp2}
\Phi (z)= \hat \phi_0 (\Lambda z)^{\epsilon} 
+(\Lambda z)^{2}\tanh{\left[(\Lambda z)^{2-\epsilon}\right]}.
\eeqa
We call this model A2. Again, the warp factor is solved numerically. The numerical solution for the warp factor in models A1 and A2 is shown  in Fig. \ref{Fig:WarpFactor} for $\epsilon=0.1$. The parameters for the model A1 (A2) are $\hat \phi_0 = 5.59$ ($5.33$) and $\Lambda=742.75 ~ {\rm MeV}$ ($677.98 ~ {\rm MeV}$). In the next subsection we will explain how those parameters were fixed.

\vspace{-0.5cm}

\subsection{The glueball spectrum}
\label{subsec:glueballs}

The dilaton-gravity equations \eqref{GenDilGravEqs} are linearised by taking $\Phi \to \Phi + \chi$ \, , \, $g_{mn} \to e^{2A(z)} \left [ \eta_{mn} + h_{mn} \right ] $ and expanding at 1\textsuperscript{st} order in $\chi$ and $h_{mn}$. The equations for these fields take the form 
\begin{align} \label{LinearDilGrav} 
R_{mn}^{(1)} &=\frac83 \partial_{(m} \Phi \,\partial_{n)} \chi -
\frac13 e^{2A} \left[ V h_{mn} +  \left(\partial_{\Phi} V\right)
 \chi \, \eta_{mn}  \right] \, , \nonumber \\
0 &= \frac43 (\nabla^2 \Phi)^{(1)} + \frac12 \left(\partial_{\Phi}^2 V\right)\chi  \, ,
\end{align} 
where  $R_{mn}^{(1)}$ and $(\nabla^2 \Phi)^{(1)}$ are the 1\textsuperscript{st} order expansions of the Ricci tensor and scalar Laplacian respectively. 

\begin{figure}[tbp]
\centering
\includegraphics[width=4.2cm,angle=0]{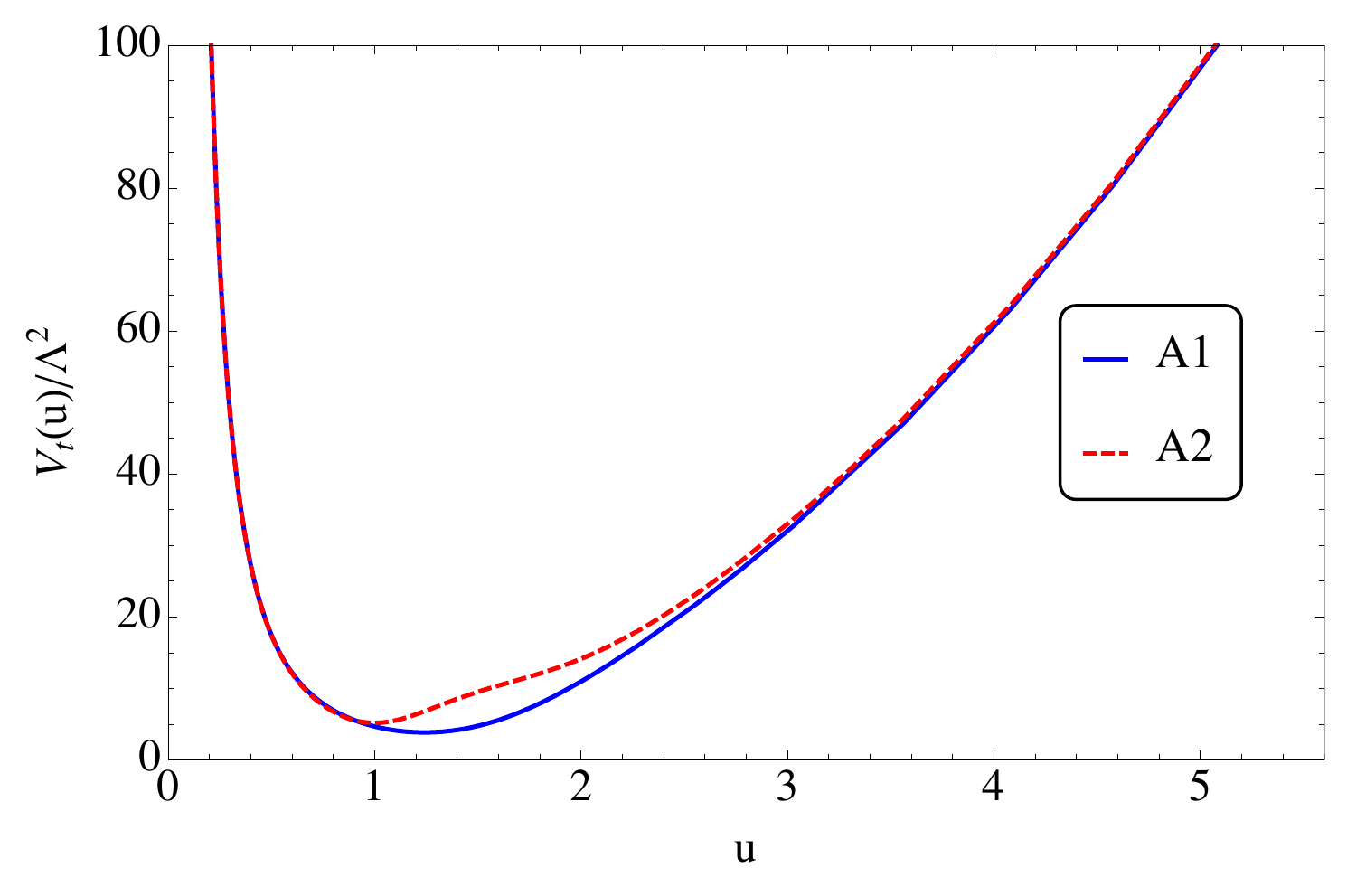} 
\includegraphics[width=4.2cm,angle=0]{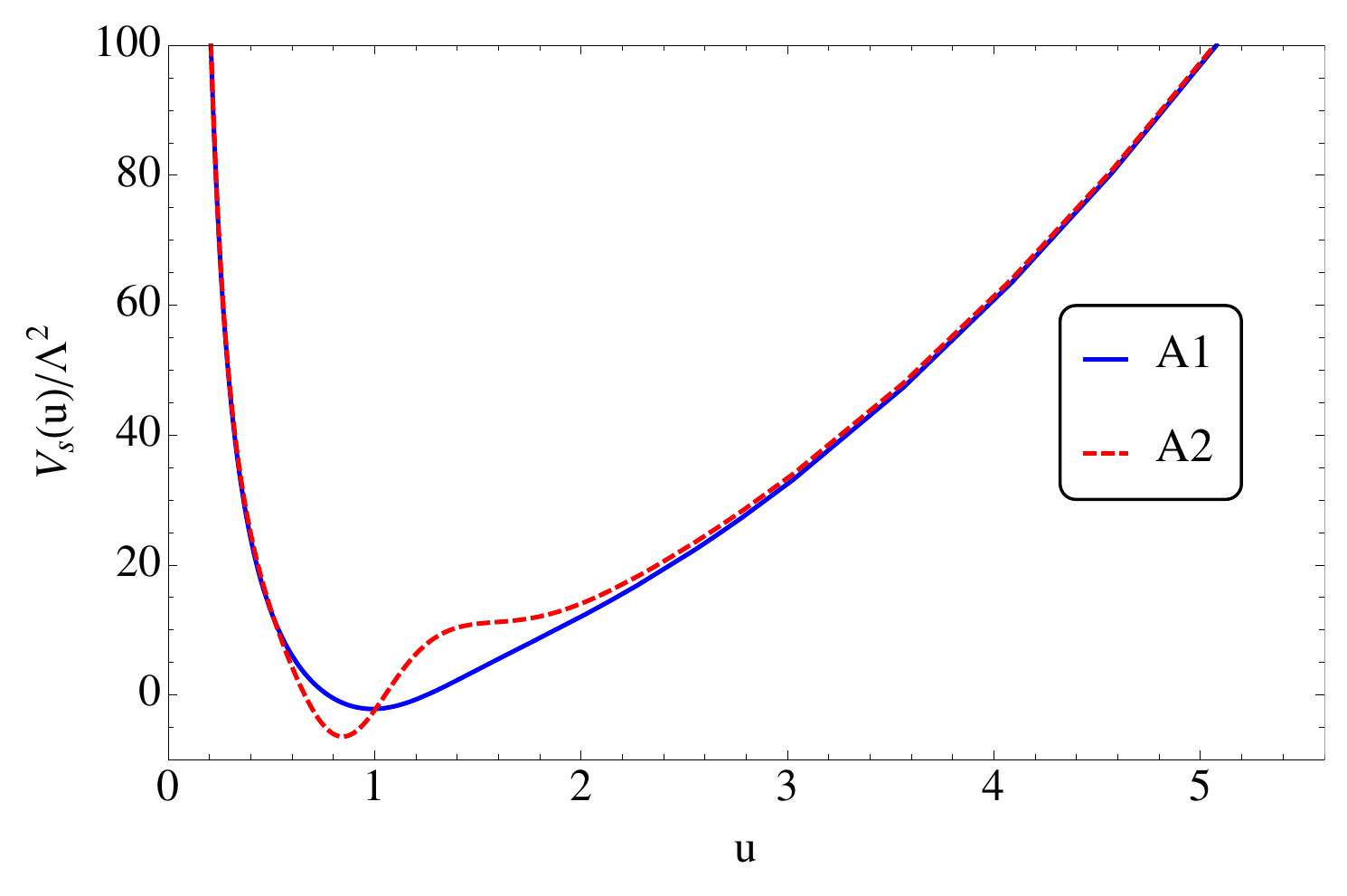}
\vspace{-0.3cm} 
\caption{The tensor (left) and scalar (right) Schr\"odinger potentials for models A1 and A2  at $\epsilon=0.01$ and  $\hat \phi_0=50$.}
\label{Fig:potentials}
\end{figure}

The metric perturbation $h_{mn}$ can be decomposed as $(h_{zz},h_{z \mu},h_{\mu \nu})$. The fields $h_{zz} \equiv 2 \phi$,  $h_{z \mu} \equiv {\cal A}_{\mu}$ and $h_{\mu \nu}$ transform as a scalar, a 4-vector and a symmetric tensor under the 4d Lorentz group respectively. Moreover, ${\cal A}_{\mu}$ and $h_{\mu \nu}$ can be further decomposed into Lorentz irreducible representations, i.e. 
\begin{align}
\mathcal{A}_{\mu} &= \mathcal{A}_{\mu}^{\scriptscriptstyle{T}} + 
\partial_{\mu} \mathcal{W}  \, , \nonumber \\
h_{\mu \nu} &= h_{\mu \nu}^{\scriptscriptstyle{TT}} + 2 \partial_{ ( \mu} 
\mathcal{V}_{\nu )}^{\scriptscriptstyle{T}}
+ 2\partial_{\mu} \partial_{\nu} \mathcal{E} + 2 \psi \eta_{\mu \nu} \,,
\label{LorentzDecomp} 
\end{align}
where $\mathcal{A}_{\mu}^{\scriptscriptstyle{T}}$ and 
$\mathcal{V}_{\mu}^{\scriptscriptstyle{T}}$ are
divergenceless vectors, $h_{\mu \nu}^{\scriptscriptstyle{TT}}$ is a traceless and divergenceless tensor and $\mathcal{W}$, $\mathcal{E}$, $\psi$ are Lorentz-scalars.  As described in \cite{Kiritsis:2006ua}, see also \cite{Ballon-Bayona:2017sxa}, one finds from \eqref{LinearDilGrav} one tensorial equation, two vectorial equations and five scalar equations. From those equations, only the following two are independent:
\begin{align}
\left [ \partial_z^2 + 3 A' \partial_z + \Box \right ] h_{\mu 
\nu}^{\scriptscriptstyle{TT}} = 0 \,, \label{Spin2eq} \\
\xi'' + \left ( 3 A' + 2 \frac{\beta_{\Phi}'}{\beta_{\Phi}} \right ) \xi' + \Box \xi &= 0 \, , 
\label{Spin0DecEq}
\end{align}
where the scalar field $\xi$ is defined by  $
\xi = \psi - \chi / \beta_{\Phi}$. The fields $h_{\mu \nu}^{\scriptscriptstyle{TT}}$ and $\xi$ describe the dynamics of the spin $2$ and spin $0$ glueballs respectively. The equations \eqref{Spin2eq} and \eqref{Spin0DecEq} can be brought into a Schr\"odinger form:
\beqa
\left [ \partial_z^2 + m_{t,s}^2 - V_{t,s} \right ] \psi_{t,s} = 0 \,, \label{Schrodeqs}
\eeqa
\begin{table}[t]
\centering
\begin{tabular}{l |c|c|c|c|c|l}
\hline 
\hline
 $n$ & A1 & A2 & B1 & B2 &
 IHQCD \cite{Gursoy:2007er}&
Lattice \cite{Meyer:2004gx} \\
\hline 
 $0^{++}$ & 1475 & 1475 & 1475 & 1475 & 1475   & 1475(30)(65)  \\
 $0^{++*}$ & 2755 & 2755 & 2755 & 2755 & 2753  & 2755(70)(120)  \\
 $0^{++**}$ & 3507 & 3376 & 3361 & 3449 & 3561  & 3370(100)(150) \\
 $0^{++***}$ & 4106 & 3891 & 3861 & 4019 & 4253  & 3990(210)(180)  \\
 $0^{++****}$ & 4621 & 4349 & 4313 & 4514 & 4860  &   \\
 $0^{++*****}$ & 5079 & 4762 & 4721 & 4956 & 5416   &  \\
 $2^{++}$ & 2075 & 2180 & 2182 & 2130  & 2055  & 2150(30)(100) \\
 $2^{++*}$ & 2945 & 2899 & 2887 & 2943 & 2991 & 2880(100)(130)  \\
 $2^{++**}$ & 3619 & 3468 & 3444 & 3568 & 3739 & \\
 $2^{++***}$ & 4185 & 3962 & 3928 & 4102  & 4396 &  \\
 $2^{++****}$ & 4680 & 4404 & 4365 & 4576  & 5530 &  \\
 $2^{++*****}$ & 5127 & 4807 & 4763 & 5006  &  &  \\
 \hline\hline
\end{tabular}
\caption{The glueball masses (MeV) in our model for $\epsilon=0.01$, compared against IHQCD \cite{Gursoy:2007er} and lattice QCD \cite{Meyer:2004gx}. The masses for $0^{++}$ and $0^{++*}$ were used as input data. }
\label{Table:spectrum}
\end{table}

with effective potentials 
\beqa
V_{t,s} = B_{t,s}'' + (B_{t,s}')^2 \, , 
\eeqa
given in terms of the functions 
\beqa
B_t = \frac32 A \, \, , \, \, B_s = \frac32 A + \log \beta_{\Phi} \,.
\eeqa
The Schr\"odinger potentials for the tensor and scalar sector are shown in Fig. \ref{Fig:potentials} as functions of the dimensionless coordinate $u = \Lambda z$. The spin $2$ and spin $0$ glueball states arise then as bound states of potential wells, obtained by solving the Schr\"odinger equations \eqref{Schrodeqs}. The corresponding eigenvalues are the squared masses of the spin $2$ and spin $0$ glueballs. We use data from lattice QCD for the first two scalar glueball masses,  $m_{0^{++}}$ and $m_{0^{++*}}$, as an input to fix $\hat \phi_0$ and $\Lambda$. For instance, when $\epsilon=0.01$ we find for the model A1 (A2) $\hat \phi_0 = 53.79$ ($49.41$) and $\Lambda = 736~{\rm MeV}$ ($682~{\rm MeV}$). The model then predicts the masses for the other glueball states. In Table \ref{Table:spectrum} we compare our results for $\epsilon=0.01$ against the results obtained in IHQCD \cite{Gursoy:2007er} and lattice QCD \cite{Meyer:2004gx}. 

\begin{figure}[tbp]
\centering
\includegraphics[width=4.2cm,angle=0]{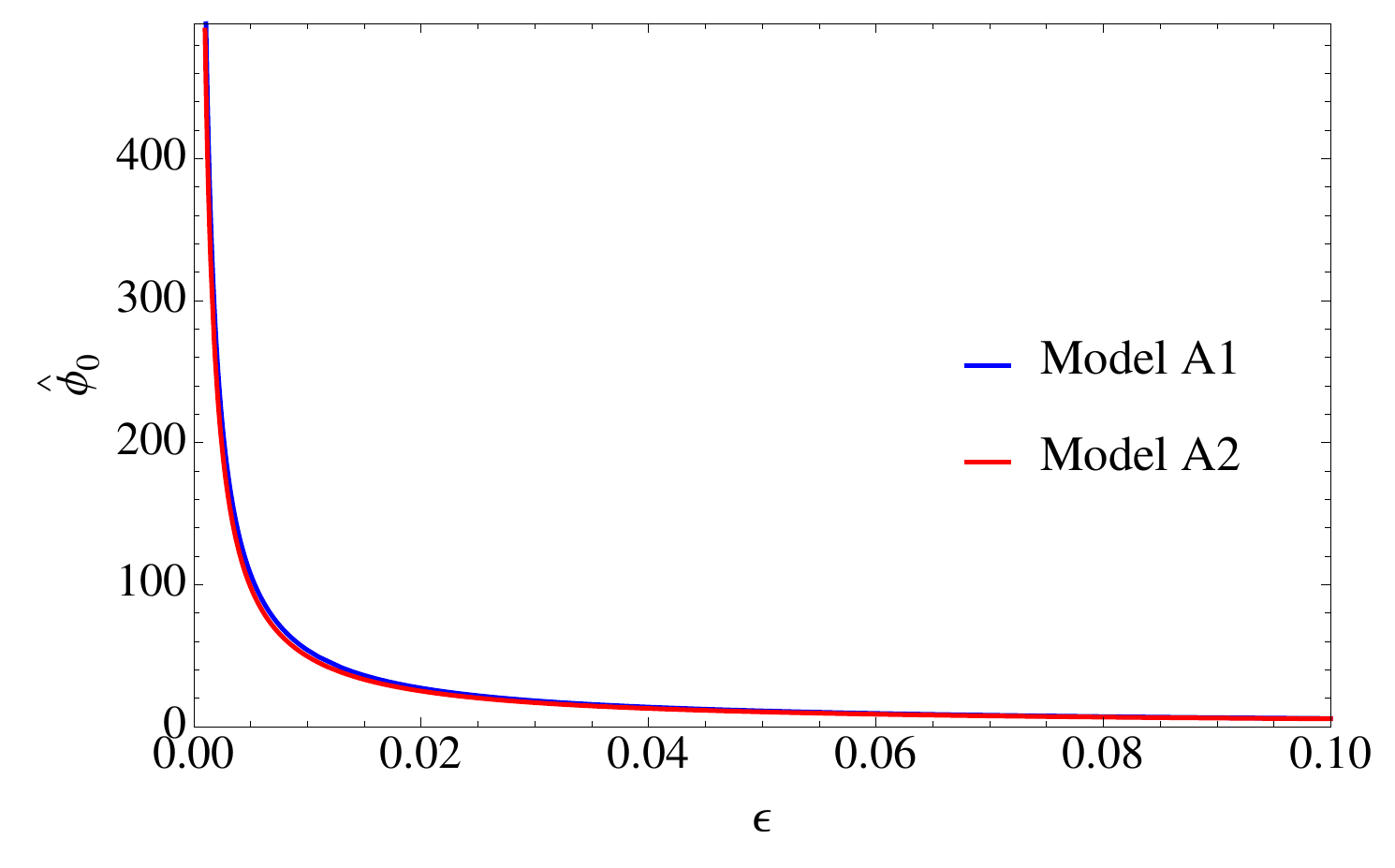} 
\includegraphics[width=4.2cm,angle=0]{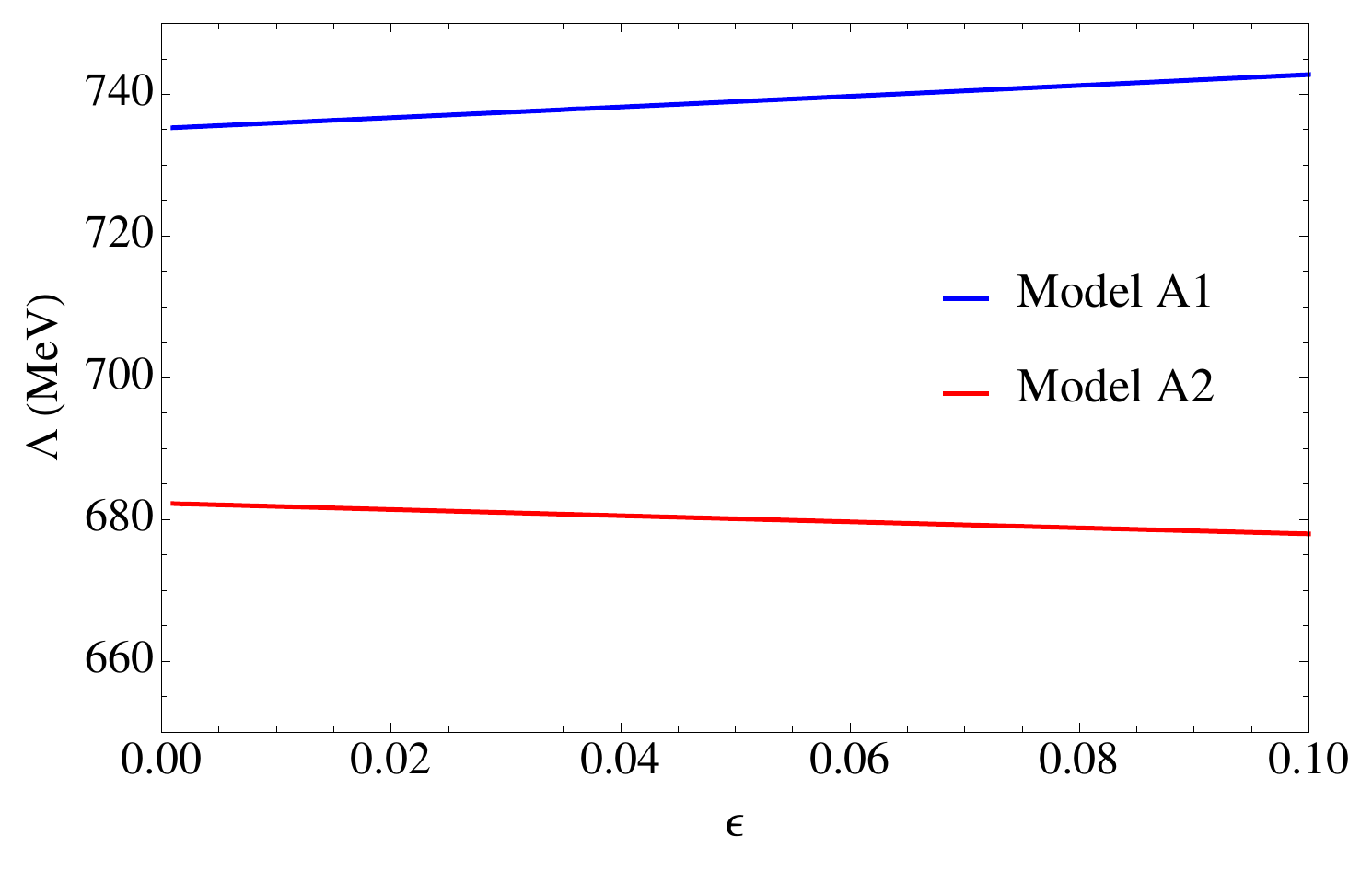}
\vspace{-0.3cm}
\caption{Evolution of the parameters $\hat \phi_0$ (left) and $\Lambda$ (right) with the conformal
dimension $\epsilon$ for models A1 and A2.}
\label{Fig:RunningParam}
\end{figure}

\vspace{-0.25cm}

\subsection{From deformed CFTs to QCD}
\label{subsec:emergbeta}

The models A1 and A2, corresponding to the interpolations  \eqref{interp1} and \eqref{interp2}, only depend on the parameters $\epsilon$, $\hat \phi_0$ and $\Lambda$. As described above, for each value of $\epsilon$ we use the masses of the first glueball states $0^{++}$ and $0^{++*}$ to fix the values of $\hat \phi_0$ and $\Lambda$. In Fig. \ref{Fig:RunningParam} we plot $\hat \phi_0$ and $\Lambda$ as functions of $\epsilon$ for the range $0.001 < \epsilon < 0.1$. The results show that the dimensionless coupling $\hat \phi_0$ 
behaves qualitatively as $1/\lambda$ with $\lambda$ being the 't Hooft coupling of large-$N_c$ QCD. On the other hand, the parameter $\Lambda$ is approximately constant. The masses of the other glueball states also remain approximately constant as $\epsilon$ varies. 

These results strongly suggest a interpretation of the deformation $\phi_0 {\cal O}$ in terms of the large-$N_c$ Yang-Mills lagrangian  ${\cal L}_{YM} = \frac{1}{\lambda}  ( \frac12 {\rm Tr} F^2 )$. In \cite{Ballon-Bayona:2017sxa} we propose the following map 
\beqa
\phi_0 = \frac{\Lambda^{\epsilon}}{\lambda} \quad , \quad 
{\cal O} = \frac12 \frac{ {\rm Tr} F^2 }{ \Lambda^{\epsilon}} \,.
\label{OpMap}
\eeqa
One advantage of the EHQCD models, is that the UV asymptotics (an $AdS$ deformation caused by a massive dilaton) belongs to the framework of holographic deformed CFTs, previously studied in the context or holographic renormalization \cite{Bianchi:2001de,Skenderis:2002wp,Papadimitriou:2016yit}. In particular, the calculation of $\langle T^{\mu \nu} \rangle $ and $ \langle {\cal O} \rangle$ is straightforward and we can reproduce the trace Ward identity \eqref{traceDefCFT}. The component $\langle T^{00} \rangle$ is interpreted as the QCD vacuum energy. On the other hand, using the map \eqref{OpMap} proposed here we can extract $\langle {\rm Tr} F^2 \rangle$ from $ \langle {\cal O} \rangle$. The result is 
\beqa
\langle {\rm Tr} \, F^2 \rangle^{ren}=
\frac{32}{15} (M_p \ell)^3 N_c^2  (4 - \epsilon) \Lambda^4  \, ,
\label{gluoncondensate}  
\eeqa
where the superscript {\it ren} means renormalized and we used a minimal subtraction (MS) scheme. Setting $(M_p \ell)^3 N_c^2$ to $1$ our results for the gluon condensate $\frac{1}{4 \pi^2} \langle {\rm Tr} F^2 \rangle$, in the $\epsilon \to 0$ limit, are  $0.063$ and $0.047\,\text{GeV}^{4}$ for models A1 and A2. Finally, using the map \eqref{OpMap} we can also rewrite the trace Ward identity \eqref{traceDefCFT} in terms of the QCD trace anomaly \eqref{QCDTraceAnom}. This leads to the following relations
\beqa
\hat \phi_0 = \frac{1}{\lambda} \quad, \quad \epsilon = \frac{- \beta_{\lambda}}{\lambda} \,.  \label{dictionary}
\eeqa
The first relation in \eqref{dictionary} is consistent with our numerical results for $\hat \phi_0$. Moreover, inverting the first plot in Fig. \ref{Fig:RunningParam} we obtain $\epsilon$ as a function of $\hat \phi_0$ and use the second relation in \eqref{dictionary} to extract the corresponding beta function $\beta_{\lambda}$. Remarkably, at small $\lambda$ the best fit of the numerical results take the form \eqref{betapQCD}, which is the same for the two-loop beta function of perturbative QCD in the large-$N_c$ limit. 

\vspace{-0.5cm}
 
\section{Conclusions}

We have described the EHQCD model proposed in \cite{Ballon-Bayona:2017sxa}, where QCD is approximated in the UV by a CFT deformed by a relevant operator of conformal dimension $4 - \epsilon$. Implementing the IR constraint found in  \cite{Gursoy:2007cb,Gursoy:2007er} we  arrive at a glueball spectrum consistent with lattice QCD data.  Analysing the evolution of the model parameters with $\epsilon$ we concluded that the CFT deformation could be identified with the Yang-Mills lagrangian. We proposed a map that allowed us to rewrite the trace Ward identity of deformed CFTs in terms of the QCD trace anomaly. The map also led to the relation $\epsilon = - \beta_{\lambda}/\lambda$ that implies that the beta function of large-$N_c$ perturbative QCD could emerge from holography.

There are some challenges that remain in the EHQCD approach. It would be desirable to find a map between the IHQCD models and the EHQCD models. That map would transform the metric near a geometric cutoff  into an $AdS$ metric. The holographic description of the Callan-Symanzik equations, following \cite{deBoer:1999tgo}, would also provide a better understanding of the EHQCD approach. Finally, extensions of this work include the description of mesons and chiral symmetry breaking, the deconfinement transition, glueball melting (following \cite{glueballmelting}), higher spin glueballs and the pomeron (following the recent progress in \cite{pomeronpapers}). 

\vspace{-0.25cm}

\section*{Acknowledgements}

This work was partially funded by Funda\c c\~ao 
de Amparo \`a Pesquisa do Estado de S\~ao Paulo (FAPESP), Brazil, Grants No.~2011/18729-1 (V. T. Z.),  No.~2013/17642-5 (L. A. H. M.) and No.~2015/17609-3 (A.B-B). V. T. Z. also thanks  Coordena\c{c}\~ao de Aperfei\c{c}oamento do Pessoal de N\'\i vel Superior (CAPES), Brazil, Grant No.~88881.064999/2014-01, and Conselho Nacional de Desenvolvimento Cient\'\i fico e Tecnol{\'o}gico  (CNPq), Brazil, Grant No.~308346/2015-7. A.B-B also 
acknowledges partial financial support from the  grant CERN/FIS-NUC/0045/2015. H. B. F. also acknowledges partial financial support from CNPq Grant No. 307278/2015-8.



\begin{thebibliography}{99}

\bibitem{Ballon-Bayona:2017sxa} 
  A.~Ballon-Bayona, H.~Boschi-Filho, L.~A.~H.~Mamani, A.~S.~Miranda and V.~T.~Zanchin,
  Phys.\ Rev.\ D {\bf 97}, no. 4, 046001 (2018)
  [arXiv:1708.08968 [hep-th]].
  
\bibitem{Edelstein:2006kw} 
  J.~D.~Edelstein and R.~Portugues,
  Fortsch.\ Phys.\  {\bf 54}, 525 (2006)
  [hep-th/0602021].
 
\bibitem{Erdmenger:2007cm} 
  J.~Erdmenger, N.~Evans, I.~Kirsch and E.~Threlfall,
  Eur.\ Phys.\ J.\ A {\bf 35}, 81 (2008)
  [arXiv:0711.4467 [hep-th]].

\bibitem{Erlich:2009me} 
  J.~Erlich,
  Int.\ J.\ Mod.\ Phys.\ A {\bf 25}, 411 (2010)
  [arXiv:0908.0312 [hep-ph]].

\bibitem{Gursoy:2010fj} 
  U.~Gursoy, E.~Kiritsis, L.~Mazzanti, G.~Michalogiorgakis and F.~Nitti,
  Lect.\ Notes Phys.\  {\bf 828}, 79 (2011)
  [arXiv:1006.5461 [hep-th]].
    
\bibitem{Brodsky:2014yha} 
  S.~J.~Brodsky, G.~F.~de Teramond, H.~G.~Dosch and J.~Erlich,
  Phys.\ Rept.\  {\bf 584}, 1 (2015)
  [arXiv:1407.8131 [hep-ph]].

\bibitem{Gursoy:2016ebw} 
  U.~Gursoy,
  Acta Phys.\ Polon.\ B {\bf 47}, 2509 (2016)
  [arXiv:1612.00899 [hep-th]].

\bibitem{Gursoy:2007cb} 
  U.~Gursoy and E.~Kiritsis,
  JHEP {\bf 0802}, 032 (2008)
  [arXiv:0707.1324 [hep-th]].
 
 \bibitem{Gursoy:2007er} 
  U.~Gursoy, E.~Kiritsis and F.~Nitti,
  JHEP {\bf 0802}, 019 (2008)
  [arXiv:0707.1349 [hep-th]].
   
\bibitem{Gubser:2008ny} 
  S.~S.~Gubser and A.~Nellore,
  Phys.\ Rev.\ D {\bf 78}, 086007 (2008)
  [arXiv:0804.0434 [hep-th]].
  
\bibitem{Gubser:2008yx}
S.~S.~Gubser, A.~Nellore, 
S.~S.~Pufu and F.~D.~Rocha, 
Phys.\ Rev.\ Lett.\  {\bf 101}, 131601 (2008)
[arXiv:0804.1950 [hep-th]].

\bibitem{Rychkov:2015naa} 
  S.~Rychkov and Z.~M.~Tan,
  J.\ Phys.\ A {\bf 48}, no. 29, 29FT01 (2015)
  [arXiv:1505.00963 [hep-th]].
  
\bibitem{Kiritsis:2006ua} 
  E.~Kiritsis and F.~Nitti,
  Nucl.\ Phys.\ B {\bf 772}, 67 (2007)
  [hep-th/0611344].
  
\bibitem{Meyer:2004gx} 
  H.~B.~Meyer,
  ``Glueball regge trajectories,'' D. Phil. Thesis, University of Oxford 
(2005) [hep-lat/0508002].

\bibitem{Bianchi:2001de} 
  M.~Bianchi, D.~Z.~Freedman and K.~Skenderis,
  JHEP {\bf 0108}, 041 (2001)
  [hep-th/0105276].
  
\bibitem{Skenderis:2002wp} 
  K.~Skenderis,
  Class.\ Quant.\ Grav.\  {\bf 19}, 5849 (2002)
  [hep-th/0209067].

\bibitem{Papadimitriou:2016yit} 
  I.~Papadimitriou,
  Springer Proc.\ Phys.\  {\bf 176}, 131 (2016).
  
\bibitem{deBoer:1999tgo} 
  J.~de Boer, E.~P.~Verlinde and H.~L.~Verlinde,
  JHEP {\bf 0008}, 003 (2000)
  [hep-th/9912012].

\bibitem{glueballmelting} 
  A.~S.~Miranda, C.~A.~Ballon Bayona, H.~Boschi-Filho and N.~R.~F.~Braga,
  JHEP {\bf 0911}, 119 (2009)
  [arXiv:0909.1790 [hep-th]].
  A.~S.~Miranda, C.~A.~Ballon Bayona, H.~Boschi-Filho and N.~R.~F.~Braga,
  Nucl.\ Phys.\ Proc.\ Suppl.\  {\bf 199}, 107 (2010)
  [arXiv:0910.4319 [hep-th]].

\bibitem{pomeronpapers} 
  
  A.~Ballon-Bayona, R.~Carcass\'es Quevedo, M.~S.~Costa and M.~Djuri\'c,
  Phys.\ Rev.\ D {\bf 93}, 035005 (2016)
  [arXiv:1508.00008 [hep-ph]].
  E.~Folco Capossoli, D.~Li and H.~Boschi-Filho,
  Phys.\ Lett.\ B {\bf 760}, 101 (2016)
  [arXiv:1601.05114 [hep-ph]].
  A.~Ballon-Bayona, R.~Carcass\'es Quevedo and M.~S.~Costa,
  JHEP {\bf 1708}, 085 (2017)
  [arXiv:1704.08280 [hep-ph]].
  
    
    
        
\end{thebibliography}
\end{document}